\documentclass[reprint,showpacs,preprintnumbers,groupedaddress,amsmath,amssymb,aps,prb]{revtex4-1}

\usepackage{graphicx}
\usepackage{dcolumn}
\usepackage{bm}

\usepackage[colorlinks]{hyperref}
\hypersetup{
pdfstartview=Fit,
pdfpagemode=UseOutlines,
pdfdisplaydoctitle=true,
pdfauthor={A. P. Sazonov et al.},
pdftitle={Field-Induced magnetic structures in Tb2Ti2O7 at low temperatures: From spin-ice to spin-flip structures.},
colorlinks=true,
citecolor={blue},
linkcolor={blue},
urlcolor={black},
breaklinks=true,
bookmarksopen=true,
bookmarksopenlevel=5,
bookmarksnumbered=true
}

\begin{document}

\title{\texorpdfstring{Field-Induced magnetic structures in Tb$_2$Ti$_2$O$_7$ at low temperatures:\\
From spin-ice to spin-flip structures.}{Field-Induced magnetic structures in Tb2Ti2O7 at low temperatures}}

\author{A. P. Sazonov}
\email{andrew.sazonov@cea.fr}
\author{A. Gukasov}
\author{I. Mirebeau}
\author{H. Cao}
\affiliation{CEA, Centre de Saclay, DSM/IRAMIS/Laboratoire L{\'e}on Brillouin, F-91191 Gif-sur-Yvette, France}
\author{P. Bonville}
\affiliation{CEA, Centre de Saclay, DSM/IRAMIS/Service de Physique de l'Etat Condens{\'e}, F-91191 Gif-Sur-Yvette, France}
\author{B. Grenier}
\affiliation{SPSMS, UMR-E 9001, CEA-INAC/UJF-Grenoble 1, MDN, F-38054 Grenoble, France}
\author{G. Dhalenne}
\affiliation{Laboratoire de Physico-Chimie de l'Etat Solide, ICMMO, Universit{\'e} Paris-Sud, F-91405 Orsay, France}

\begin{abstract}
We studied the field-induced magnetic structures of the Tb$_2$Ti$_2$O$_7$ pyrochlore by single-crystal neutron diffraction with a magnetic field applied along a [110] axis, focusing on the influence of a small misalignment. Both induced magnetic structures with $\bm{k} = \mathbf{0}$ and $\bm{k} = (0,0,1)$ propagation vectors are found to be sensitive to the misalignment, which controls the magnitude and orientation of the Tb moments involved in the $\beta$ chains, with local [111] anisotropy axis perpendicular to the field. For $\bm{k} = \mathbf{0}$, spin-ice-like structures are observed for a misalignment of a few degrees, whereas other structures, where the Tb-$\beta$ moments flip by ``melting'' on the field axis, occur when the field is perfectly aligned. The field evolution of the $\bm{k} = \mathbf{0}$ structure is well reproduced by a molecular field model with anisotropic exchange. We give a complete symmetry analysis of the $\bm{k} = \mathbf{0}$ and $\bm{k} = (0,0,1)$ magnetic structures, both being described by the basis functions of single irreducible representations.
\end{abstract}


\maketitle

\section{Introduction}

Geometrically frustrated magnets such as spin liquids and spin ices bring out fundamental questions about their ground state and allow low-energy spin excitations to be observed. Historically~\cite{zfp.33.31.1979}, a spin liquid is defined as a cooperative paramagnet, where spins fluctuate down to zero temperature with spin-spin correlations rapidly decaying with increasing interatomic distances. From theory, it is now well established that the antiferromagnetic (AFM) interactions between isotropic (Heisenberg) spins yield a spin-liquid ground state for peculiar geometries, like the kagome, checkerboard, and pyrochlore lattices~\cite{zfp.33.31.1979,cjp.79.1323.2001,jpcm.16.S759.2004}. Quantum spin-liquid states have also been found in dimer models~\cite{prl.61.2376.1988}, when the ground state is described by a superposition of spin singlets without symmetry breaking. The characteristic of spin liquids is the continuous degeneracy of the ground-state manifold, leading to macroscopic entropy at $T = 0$\,K. By contrast, spin ices refer to systems with discrete degeneracy issuing from topological constraints, like those which govern the two energetically equivalent positions of protons of water molecules in real ice. Hence, spin ices have the same residual entropy as water ice, and they have been found in kagome and pyrochlore lattices, where ferromagnetically coupled Ising spins are constrained to lie along their local anisotropy axes~\cite{sc.294.1495.2001,rmp.82.53.2010}. Such constraints yield exotic excitations akin to magnetic monopoles~\cite{nt.451.42.2008}, which locally violate the ice rules by stochastic propagations along Dirac strings.

Rare-earth titanate pyrochlores $R_2$Ti$_2$O$_7$, where the rare-earth ions $R$ belong to a lattice of corner sharing tetrahedra, show well-known examples of spin-liquid ($R =$ Tb) and spin-ice ($R =$ Dy, Ho) behaviors. Tb$_2$Ti$_2$O$_7$ is especially interesting since the origin of its spin-liquid ground state~\cite{prl.82.1012.1999} remained mysterious for many years. Simple considerations would predict a transition toward a nonfrustrated N{\'e}el order with Ising local anisotropy, in contrast to experiment. Indeed, Tb$_2$Ti$_2$O$_7$ shows dynamic short-range correlations, without onset of long-range magnetic order down to the lowest measured temperature of 50\,mK. The energy terms responsible for this anomaly could be the peculiar crystal field scheme of the Tb$^{3+}$ ion, with a low-energy splitting between the ground and first excited doublets~\cite{prb.62.6496.2000,prb.76.184436.2007}, and an effective Tb-Tb first neighbor interaction at the verge between AFM and ferromagnetic. Recent theories suggest that crystal field fluctuations turn the effective Tb-Tb interaction to ferromagnetic, leading to a fluctuating quantum spin-ice state~\cite{prl.98.157204.2007}. Tb$_2$Ti$_2$O$_7$ may therefore represent the ``missing link'' between spin liquid and spin ice, having a highly tunable ground state, easily destabilized by magnetic field~\cite{prl.96.177201.2006}, pressure~\cite{nt.420.54.2002,prl.93.187204.2004}, or chemical substitution~\cite{prl.94.246402.2005}.

In Tb$_2$Ti$_2$O$_7$ magnetic long-range order can be induced by a magnetic field applied along the [110] direction~\cite{prl.96.177201.2006,prl.101.196402.2008}. The field-induced structures are of two types, with $\bm{k} = \mathbf{0}$ and $\bm{k} = (0,0,1)$ propagation vectors. The $\bm{k} = \mathbf{0}$ structure has non-zero magnetization, whereas the $\bm{k} = (0,0,1)$ one is antiferromagnetic. In our previous work~\cite{prl.101.196402.2008,jpcs.145.012021.2009}, we studied their local spin configurations by neutron diffraction versus field and temperature. In the $\bm{k} = \mathbf{0}$ structure, we found that the local order in low fields and at low temperatures is akin to a spin ice, but with different values of the magnetic moments in a tetrahedron. The $\bm{k} = (0,0,1)$ structure was observed with a finite length scale, in a limited range of the phase diagram above 2\,T and below 2\,K, coexisting with the $\bm{k} = \mathbf{0}$ one.

Recent measurements in Ho$_2$Ti$_2$O$_7$ spin ice have shown that the field-induced structures are very sensitive to the precise alignment of the field with respect to the crystal axes~\cite{prb.79.014408.2009}. This effect was qualitatively observed by checking the field dependence of some magnetic Bragg peaks, and it was even used as a trick to tune the concentration of magnetic monopoles or the dimer concentration in the Kasteleyn state~\cite{np.3.566.2007,sc.326.411.2009,jpsj.78.103706.2009}. It results from the huge uniaxial anisotropy of the paramagnetic susceptibility in spin ices~\cite{prl.103.056402.2009}. Considering the uniaxial anisotropy, when $\bm{H}$ is along the [110] axis, the field induced structures involve the so-called $\alpha$ and $\beta$ chains~\cite{jpsj.73.1619.2004,prl.101.196402.2008} along perpendicular directions. The $\alpha$ chains have their local anisotropy axis at 36$^\circ$ from the field whereas the $\beta$ chains have their easy axis perpendicular to $\bm{H}$. In Ho$_2$Ti$_2$O$_7$, the $\bm{k} = (0,0,1)$ structure was observed only when the field was perfectly aligned along the [110] crystal axis~\cite{prl.79.2554.1997,prb.79.014408.2009}.

Here, we have checked the influence of a field misalignment in Tb$_2$Ti$_2$O$_7$. By varying the field alignment with respect to a [110] axis in a systematic way, and refining the magnetic structures in each case, we determine how the local spin arrangements are affected. The misalignment mainly influences the moments on the Tb-$\beta$ chains and the balance between the two structures. The $\bm{k} = (0,0,1)$ structure, which involves almost only the Tb-$\beta$ moments, is favored when the alignment is optimal, and disappears when it exceeds 4.5$^\circ$. This explains previous discrepancies concerning the transition line in the magnetic phase diagram of Tb$_2$Ti$_2$O$_7$. In the $\bm{k} = \mathbf{0}$ structure, the magnitude and orientation of the Tb-$\beta$ moments also strongly depend on the misalignment. The spin-ice orders observed in low fields are actually stable only when the field is slightly misaligned. An optimal alignment favors Tb-$\beta$ moments opposite to the field, up to a compensation field of 1.5(5)\,T where they vanish, and above which they start to grow  parallel to the field. All these features are quantitatively explained using a four sites molecular field model with anisotropic exchange.

We performed a symmetry analysis, taking into account the symmetry lowering with respect to the $Fd\bar{3}m$ space group induced by a field applied along [110]. We find that both $\bm{k} = \mathbf{0}$ and $\bm{k} = (0,0,1)$ structures can be described by irreducible representations in the space group $I4_1/amd$, the highest subgroup of $Fd\bar{3}m$ allowing the presence of a homogeneous magnetization $M_z$ along a twofold axis, invariant by symmetry. The symmetry analysis describes the magnetic structures with a limited number of parameters, including the case of a misoriented field. It provides a sound basis for the description of the pyrochlore lattice with [110] magnetic field by means of the $\alpha$ and $\beta$ chains, which can be useful for real spin ices.

\section{Experiment}

A single crystal of Tb$_2$Ti$_2$O$_7$ was grown from a sintered rod of the same nominal composition by the floating-zone technique, using a mirror furnace~\cite{prl.101.196402.2008}. Neutron-diffraction studies were performed on the diffractometer 5C1~($\lambda = 0.845$\,\AA) at the Orph{\'e}e reactor of the Laboratoire L{\'e}on Brillouin, Saclay and on the CRG-CEA diffractometer D23~($\lambda=1.2815$\,\AA) at the Institut Laue-Langevin, Grenoble. We used unpolarized neutrons, collecting typically from 200 to 400 reflections for each data set. We collected data under external field applied close to a [110] direction. The best orientation we achieved corresponds to a field misaligned from [110] by $\Delta\phi = 0.4$$^\circ$ in the (001) plane to within an accuracy of $\pm 0.3$$^\circ$. The field misalignment angle as well as the precision of the alignment were estimated based on the least-squares refinement of the orientation matrix using 20 reflections measured several times. Then the field was progressively inclined by manual adjustment of the goniometer head from the [110] to the $[1\bar{1}0]$ direction thus always remaining in the (001) plane. The field misalignment angle $\Delta\phi$ was varied between $-3$$^\circ$ and $+16$$^\circ$, the temperature $T$ between 0.1\,K and 4\,K and the field $H$ between 0\,T and 12\,T. The program \textsc{FullProf} (Ref.~\onlinecite{phb.192.55.1993}) was used to refine the crystal and magnetic structure parameters.

Prior to the low-temperature studies with applied field the crystal was characterized in zero field. A total of 289 reflections with $\ensuremath{\sin\theta/\lambda} < 0.55$\,\AA$^{-1}$ were measured at 1.6\,K and 44 unique reflections were obtained by averaging equivalents ($R_\mathrm{int} = 0.043$) using the cubic space group $Fd\bar{3}m$. These were used to refine one oxygen positional parameter, all the isotropic temperature factors, the scale and the extinction parameters ($R_\mathrm{F} = 0.037$).

\section{Results and discussion}

\subsection{\label{s:NeutrDiff}Influence of the field misalignment on the phase diagram}

\begin{figure}
\includegraphics[width=0.986\columnwidth]{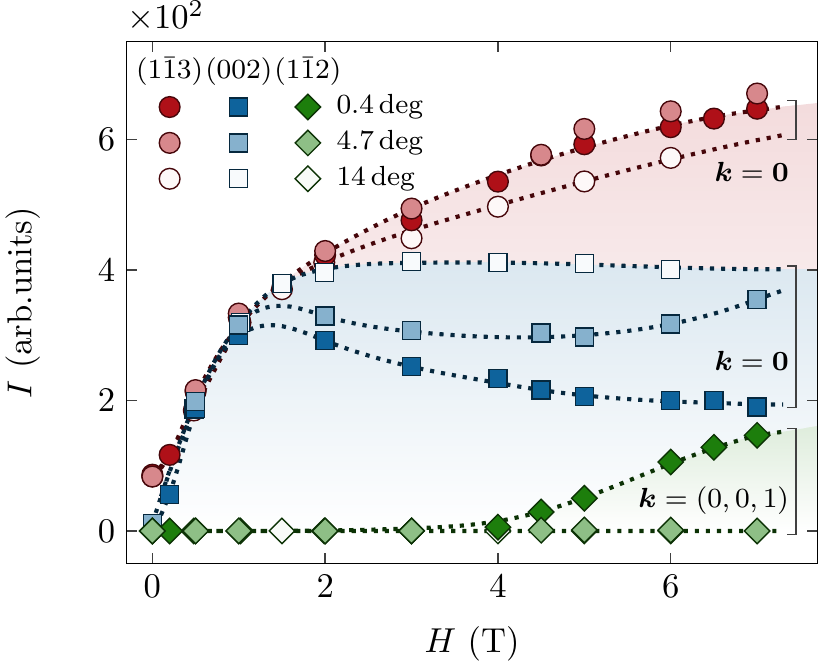}
\caption{\label{f:IvsH}(Color online) Field dependences of selected Bragg reflections of both magnetic structures with $\bm{k} = \mathbf{0}$ and $\bm{k} = (0,0,1)$ at 1.6\,K and different field misalignment angles $\Delta\phi$ with respect to the [110] direction. Error bars are smaller than the symbol size if not given. The dotted lines are guide to the eyes.}
\end{figure}

\begin{figure}
\includegraphics[width=1.0\columnwidth]{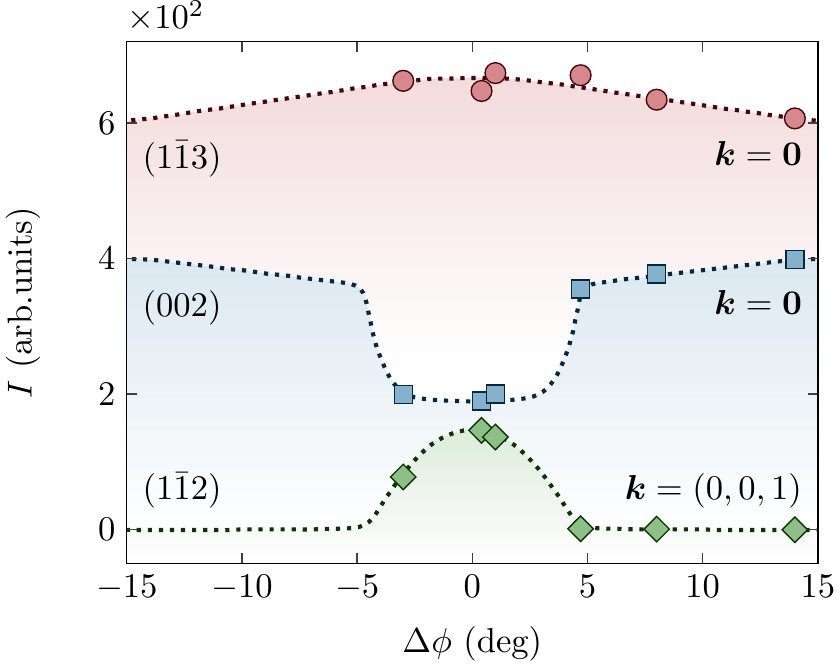}
\caption{\label{f:IvsAngle}(Color online) Intensities of selected reflections measured at 1.6\,K and 7\,T as a function of the field misalignment angle $\Delta\phi$. Error bars are smaller than the symbol size if not given. The dotted lines are guide to the eyes.}
\end{figure}

To determine the conditions of  stability of the $\bm{k} = (0,0,1)$ structure, three Bragg reflections of different types were considered. The first reflection ($1\bar{1}$3) has both nuclear and $\bm{k} = \mathbf{0}$ magnetic components. The other two reflections (002) and ($1\bar{1}$2) are purely magnetic and belong to the $\bm{k} = \mathbf{0}$ and $\bm{k} = (0,0,1)$ structures, respectively. They appear only under an applied magnetic field. In Fig.~\ref{f:IvsH},
%
the intensities of these reflections are plotted at 1.6\,K versus the magnetic field for different angles between the field and the [110] direction. The intensity of the ($1\bar{1}$3) peak is only weakly affected by the misalignment. In contrast, the intensities of the (002) and ($1\bar{1}$2) Bragg peaks strongly vary in opposite ways. The ($1\bar{1}$2) peak which has about the same intensity as the (002) peak for a well-aligned sample at 7\,T, is completely suppressed for a misalignment above 4.5$^\circ$, whereas the intensity of the (002) peak increases by about a factor of 2. Similar observations can be made from Fig.~\ref{f:IvsAngle},
%
where the intensities of the three Bragg peaks at 1.6\,K and 7\,T are plotted versus the field misalignment angle $\Delta\phi$.

\begin{figure}\vspace{2.3ex}
\includegraphics[width=1.0\columnwidth]{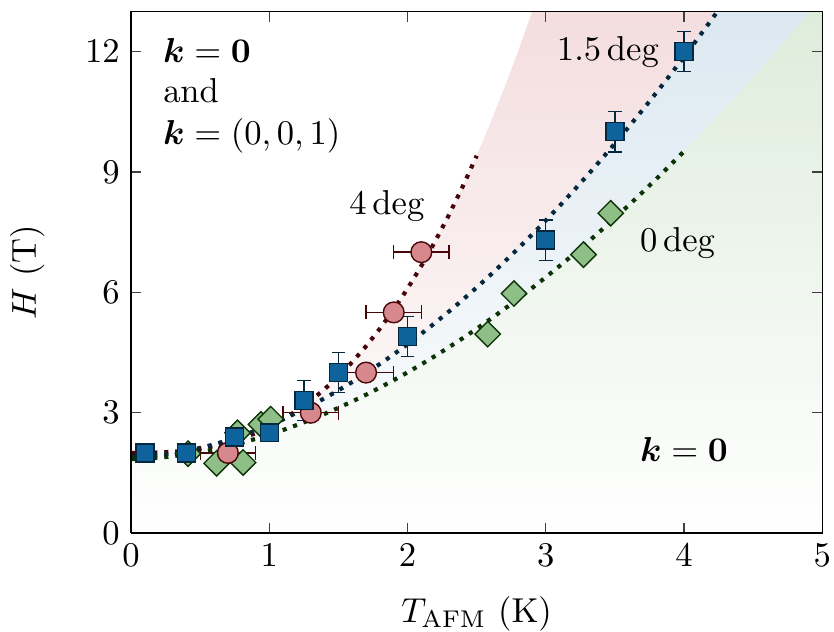}
\caption{\label{f:HvsT}(Color online) $H$--$T_\mathrm{AFM}$ phase diagram at different field misalignment angles with respect to the [110] direction according to our present measurements (squares), Ref.~\onlinecite{prl.101.196402.2008} (circles) and Ref.~\onlinecite{prl.96.177201.2006} (diamonds). The typical alignment error is less than 0.5$^\circ$. The dotted lines are guide to the eyes.}
\end{figure}

The suppression of the ($1\bar{1}$2) peak and of all other peaks of the same family shows the vanishing of the $\bm{k} = (0,0,1)$ structure. By measuring the temperature and field dependences of such peaks for a given field misalignment, we have drawn in Fig.~\ref{f:HvsT}
%
the transition lines in the $H$--$T_\mathrm{AFM}$ phase diagram of Tb$_2$Ti$_2$O$_7$. For a misalignment of 1.5$^\circ$, the $\bm{k} = (0,0,1)$ structure persists up to the highest measured field, and its transition temperature increases with the field, reaching 4\,K at 12\,T. Above 4.5$^\circ$ misalignment, no $\bm{k} = (0,0,1)$ structure is observed. Comparison with previous results with other field misalignments from Refs.~\onlinecite{prl.96.177201.2006,prl.101.196402.2008} shows that the onset of the $\bm{k} = (0,0,1)$ structure always occurs above a critical field of 2\,T but that its stability range strongly depends on the misalignment. For a given field, the transition temperature decreases with increasing the misalignment. This explains the difference between the transition lines in $H$--$T_\mathrm{AFM}$ phase diagram of Tb$_2$Ti$_2$O$_7$ previously reported in the literature~\cite{prl.96.177201.2006,prl.101.196402.2008}.

In the same way, the reentrant character of the $\bm{k} = (0,0,1)$ structure quoted in Refs.~\onlinecite{prl.101.196402.2008,jpcs.145.012021.2009}, can now be attributed to the influence of the field misalignment. In these papers, the $\bm{k} = (0,0,1)$ structure was found to be stabilized below 1\,K in a limited field range, with an onset at 2\,T and a maximum of the intensity of the AFM peaks and correlation length around 5\,T. Therefore, it was suggested that the $\bm{k} = (0,0,1)$ structure would disappear at high fields. The present measurements performed at 0.3\,K up to very high fields show that this structure indeed occurs at 2\,T, but remains stable at least up to 12\,T when the sample is well aligned.

From the opposite variations in the (002) and ($1\bar{1}$2) Bragg peaks, we also conclude that the balance between $\bm{k} = \mathbf{0}$ and $\bm{k} = (0,0,1)$ structures is affected by the field misalignment. Magnetic refinement and symmetry analysis, performed separately for each structure, yield a quantitative characterization of this interplay.

\subsection{\label{s:SymmMagn}\texorpdfstring{Symmetry analysis of the $\mathbf{k=0}$ structures}{Symmetry analysis of the k=0 structures}}

To analyze the field-induced magnetic structure of Tb$_2$Ti$_2$O$_7$, a systematic search was performed based on the theory of representations of space groups proposed by Bertaut~\cite{acra.24.217.1968} and Izyumov~\cite{jmmm.12.239.1979}. It allows one to consider all possible models of magnetic structures consistent with a given crystal structure of space group $G$. According to this method the magnetic structure can be expressed via the basis vectors of the irreducible representations of the group $G$. To calculate the irreducible representations the program BasIreps from the \textsc{FullProf} suite~\cite{phb.192.55.1993} was used.

\begin{figure*}
\includegraphics[width=0.9\textwidth]{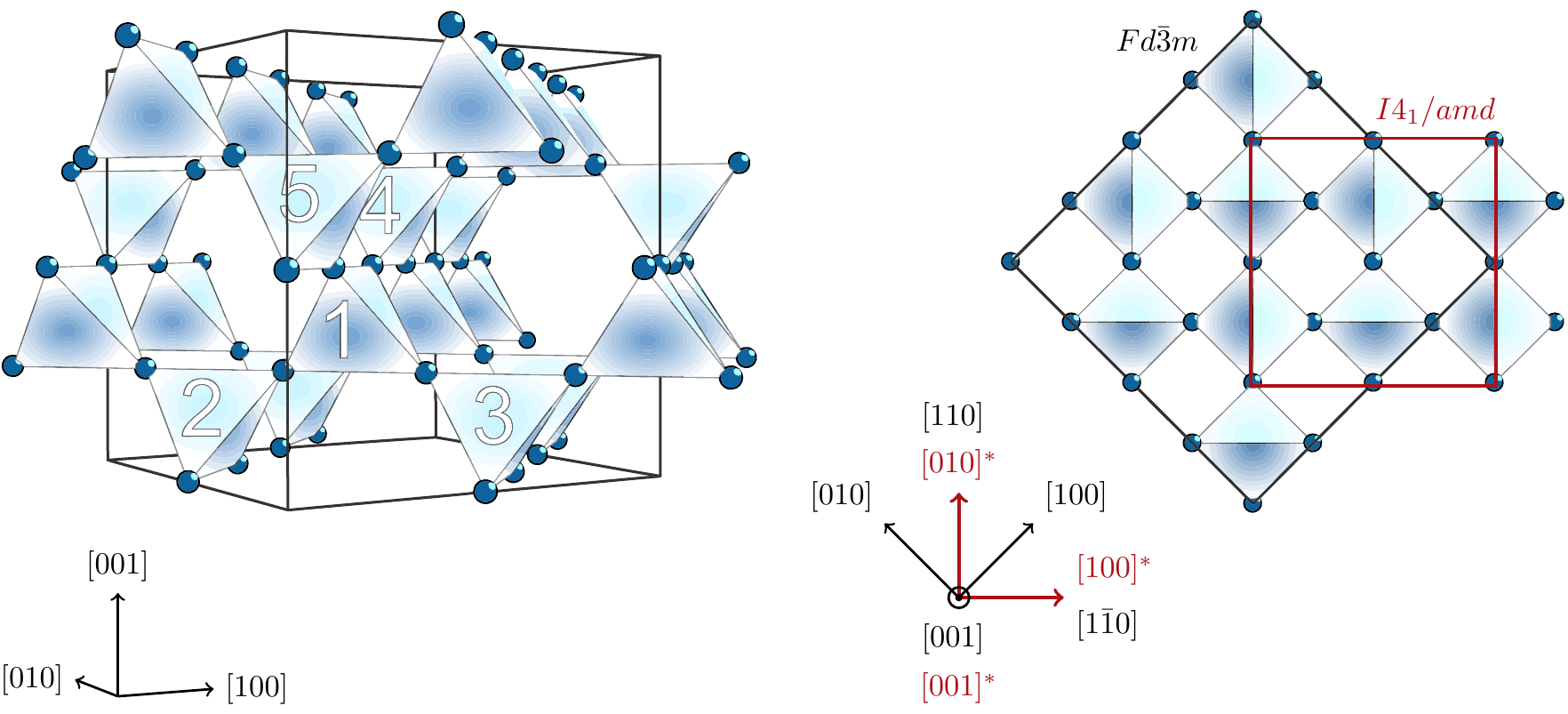}
\caption{\label{f:CrystalStructure}(Color online) Left: crystal structure of Tb$_2$Ti$_2$O$_7$. Only the Tb ions are shown for simplicity. Some tetrahedra are numbered for an easier comparison with Figs.~\ref{f:MagnStrA}, \ref{f:MagnStrB}, \ref{f:MagnStrC} and~\ref{f:MagnStrD}. Right: transformation from the $Fd\bar{3}m$ to $I4_1/amd$ space group; view along the [001] direction.}
\end{figure*}

In zero field there are four irreducible representations for the space group $Fd\bar{3}m$ associated with the propagation vector $\bm{k} = \mathbf{0}$. However, none of the zero-field magnetic structures allowed by the $Fd\bar{3}m$ symmetry are compatible with the experimental data obtained under field $\bm{H} \parallel [110]$. This suggests that the magnetic field applied along the twofold [110] axis breaks the $Fd\bar{3}m$ symmetry. Therefore, we searched for new solutions in the space group $I4_1/amd$, the highest subgroup of $Fd\bar{3}m$, for which a homogeneous magnetization component $M_z$ induced by the field applied along the two-fold axis is invariant.

The crystal structure of $R_2$Ti$_2$O$_7$ pyrochlore compounds in the $Fd\bar{3}m$ space group is shown in Fig.~\ref{f:CrystalStructure} together with the transformation from the $Fd\bar{3}m$ to the $I4_1/amd$ space group. We notice that although a distortion of the cubic structure was inferred in Tb$_2$Ti$_2$O$_7$ below 20\,K from x-ray measurements~\cite{prl.99.237202.2007}, it remains much too small to be observed with the resolution of neutron diffraction. The new unit-cell parameters in $I4_1/amd$ are $a^*=a\frac{\sqrt{2}}{2}$ and $c^*=a$, where $a$ is the cubic unit-cell parameter of the $Fd\bar{3}m$ space group. Hereafter, the symbol~* will be used to denote the lattice parameters, coordinates and directions associated with the $I4_1/amd$ symmetry. The positions of the four Tb sites (in $I4_1/amd$) are defined in Table~\ref{t:IrrRepTb1}. They are labeled according to the $\alpha$ (Tb-$\alpha_1$ and Tb-$\alpha_2$) or $\beta$ (Tb-$\beta_1$ and Tb-$\beta_2$) chains to which they belong.

We start by considering that, in spite of the fact that the $\alpha$ and $\beta$ chains can have different moments, they are both involved in the phase transition with $\bm{k} = \mathbf{0}$. In this case it is generally accepted that their basis functions should belong to the same irreducible representation. The magnetic representation for the Tb$^{3+}$ ions occupying the $8d$ site in the $I4_1/amd$ space group can be written as $\Gamma_\mathrm{Tb} = 1\,\Gamma^{(1)}_{1} + 1\,\Gamma^{(1)}_{3} + 2\,\Gamma^{(1)}_{6} + 2\,\Gamma^{(1)}_{8} + 3\,\Gamma^{(2)}_{9}$, following the numbering scheme of Kovalev~\cite{book.kovalev.1965}, where the superscript is the order of the irreducible representation. The basis vectors describing the Tb moments which transform according to the irreducible representations $\Gamma_1$, $\Gamma_3$, $\Gamma_6$ and $\Gamma_8$ can be ruled out on the basis of magnetization measurements as none of them allows the existence of a net ferromagnetic moment along the field direction [010]$^*$. In contrast, the two-dimensional representation $\Gamma_{9}$ described in Table~\ref{t:IrrRepTb1} allows the presence of both a ferromagnetic order along the [010]$^*$ direction (see the $M_y$ component in Table~\ref{t:IrrRepTb1}) and an antiferromagnetic order along the [001]$^*$ direction (see the $M_z$ component). Moreover, $\Gamma_{9}$ also allows a net magnetization component along [100]$^*$ (see the $M_x$ component). As a result, the magnetic structure induced by a field misaligned from the [010]$^*$ direction but remaining in the (001)$^*$ plane can also be described by the same irreducible representation $\Gamma_{9}$.

We note that in the basis functions of $\Gamma_{9}$, the magnetic moments in the Tb-$\alpha$ and $\beta$ chains are completely decoupled and can be varied separately. To summarize, for the refinement of the $\bm{k} = \mathbf{0}$ structure one can use three fitting parameters $U_x$, $U_y$, and $U_z$ for the Tb-$\alpha$ chains and three other parameters $V_x$, $V_y$, and $V_z$ for the Tb-$\beta$ chains. Therefore, the symmetry analysis reduces the number of the refined parameters from twelve in an unconstrained refinement (three for each of the four Tb moments in a tetrahedron), to six only.

\begin{table}
\caption{\label{t:IrrRepTb1} Irreducible representation $\Gamma_{9}$ of Tb$_2$Ti$_2$O$_7$ (space group $I4_1/amd$) associated with $\bm{k} = \mathbf{0}$. Basis vectors projected from a general vector $\bm{M}$ with the components $M_x$, $M_y$, and $M_z$ at the Tb $8d$ sites. Tb-$\alpha_1$: $(x,y,z)^*$, Tb-$\alpha_2$: $(-x+\frac{1}{2},-y,z+\frac{1}{2})^*$, Tb-$\beta_1$: $(y+\frac{1}{4},x+\frac{3}{4},-z+\frac{1}{4})^*$ and Tb-$\beta_2$: $(-y+\frac{1}{4},-x+\frac{1}{4},-z+\frac{3}{4})^*$.}
\begin{ruledtabular}
\begin{tabular}{llrrr}
Irr. Rep.     &Atom          &$M_x$  &$M_y$  & $M_z$  \\
\colrule
$\Gamma_{9}$  &Tb-$\alpha_1$ &$U_x$  &$U_y$  & $U_z$  \\
              &Tb-$\alpha_2$ &$U_x$  &$U_y$  &$-U_z$  \\
              &Tb-$\beta_1$  &$V_x$  &$V_y$  &$-V_z$  \\
              &Tb-$\beta_2$  &$V_x$  &$V_y$  & $V_z$  \\
\end{tabular}
\end{ruledtabular}
\end{table}

\subsection{\label{s:MagnStr}\texorpdfstring{Magnetic refinements of the $\mathbf{k=0}$ structure}{Magnetic refinements of the k=0 structure}}

\begin{figure*}
\includegraphics[width=1.0\textwidth]{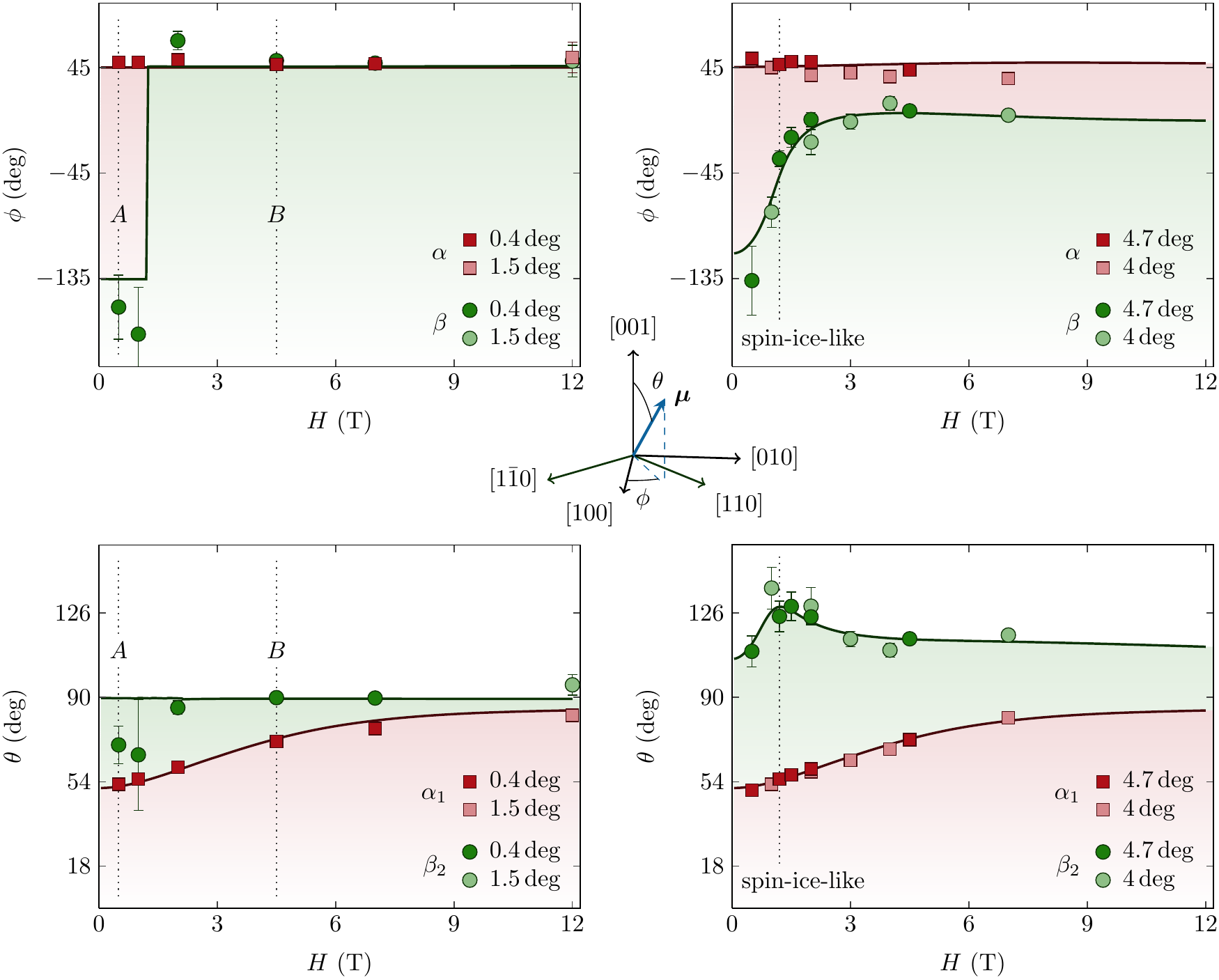}
\caption{\label{f:MagnAngle}(Color online) Tb$_2$Ti$_2$O$_7$ $\bm{k} = \mathbf{0}$ structure at 1.6\,K. Field dependences of the Tb-$\alpha$ and $\beta$ moment angles for different field misalignments ($\Delta\phi$) with respect to the [110] direction. Error bars are smaller than the symbol size if not given. The angles $\phi$ (top) and $\theta$ (bottom) of a given moment $\mu$ are defined at the center in spherical coordinates. The $\alpha_2$ ($\beta_2$) moment is symmetric of $\alpha_1$ ($\beta_1$) with respect to the (001) plane. Left: well-oriented samples with $\Delta\phi = 0.4$$^\circ$ and 1.5$^\circ$. Right: misoriented samples with $\Delta\phi = 4$$^\circ$ and 4.7$^\circ$. The solid lines are calculations with the model described in Sec.~\ref{s:Model}. The following parameters were used for the calculations: $\lambda_\perp = -0.42$\,T/$\mu_\mathrm{B}$, $\lambda_\parallel  = -0.06$\,T/$\mu_\mathrm{B}$, $\Delta\phi = 0$$^\circ$ (left) and $\Delta\phi = 3$$^\circ$ (right). The magnetic structures corresponding to the angles marked with ``$A$'', ``$B$'' and ``spin-ice-like'' (vertical dotted lines) are shown in Figs.~\ref{f:MagnStrA} and~\ref{f:MagnStrB} and described in the text.}
\end{figure*}

To refine the $\bm{k} = \mathbf{0}$ structure of Tb$_2$Ti$_2$O$_7$, from 200 to 400 Bragg reflections were collected up to $\ensuremath{\sin\theta/\lambda} \approx 0.55$\,\AA$^{-1}$. Such a collection was made in each case, namely, for given values of the magnetic field, temperature and field misalignment. The integrated intensities of these reflections were used to refine the components of the Tb$^{3+}$ magnetic moments using the symmetry constraints described above. We note that with this reduced number of parameters, we obtain the same or even better quality of refinement as in previous unconstrained refinements~\cite{prl.101.196402.2008,jpcs.145.012021.2009}. Nuclear structure parameters were taken from the low-temperature measurements in zero field. We found that, among all possible irreducible representations for $\bm{k} = \mathbf{0}$, only $\Gamma_{9}$ (see Table~\ref{t:IrrRepTb1}) provides a reliable description of the neutron-diffraction data.

\begin{figure}
\includegraphics[width=1.0\columnwidth]{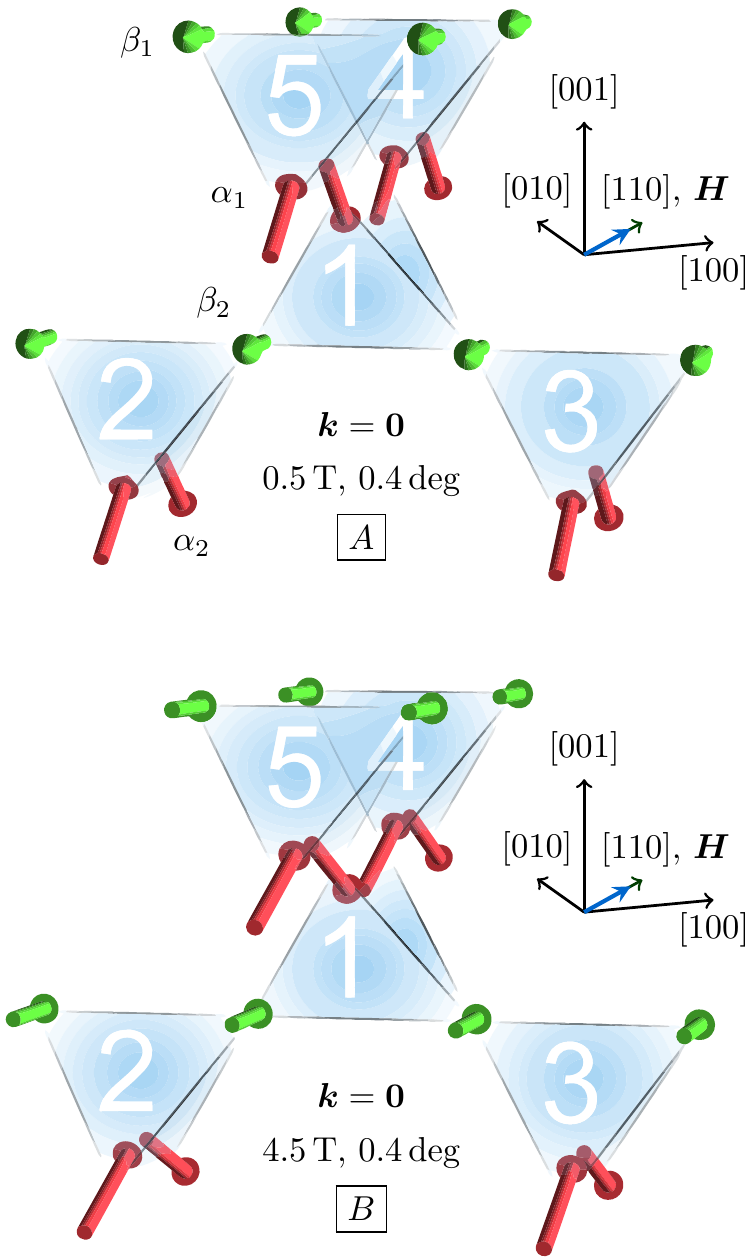}
\caption{\label{f:MagnStrA}(Color online) Field-induced magnetic structure with $\bm{k} = \mathbf{0}$ of Tb$_2$Ti$_2$O$_7$ in the well-aligned sample ($\Delta\phi = 0.4$$^\circ$) below ($A$) and above ($B$) the flip of the Tb-$\beta$ moments. Tetrahedra are numbered for an easy comparison with Fig.~\ref{f:CrystalStructure}. Magnetic moment angles corresponding to structures denoted by $A$ and $B$ are shown in Fig.~\ref{f:MagnAngle} and described in the text.}
\end{figure}

\begin{figure}
\includegraphics[width=1.0\columnwidth]{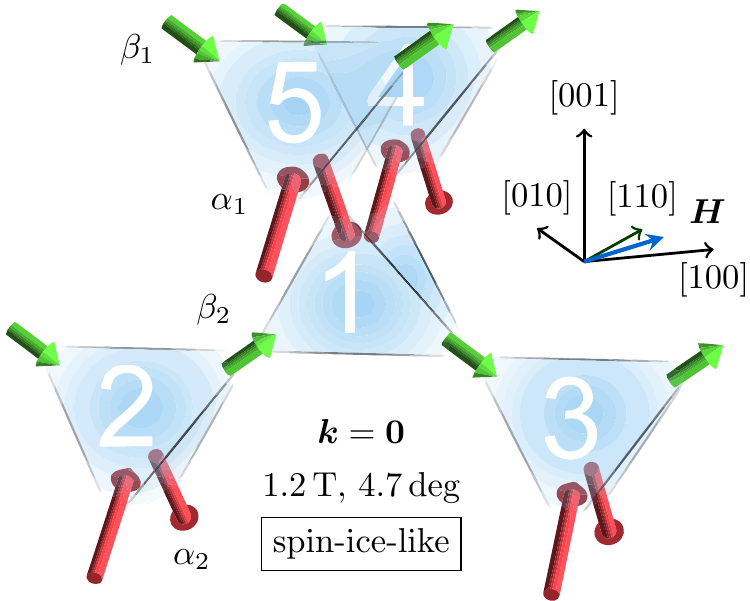}
\caption{\label{f:MagnStrB}(Color online) Field-induced magnetic structure with $\bm{k} = \mathbf{0}$ of Tb$_2$Ti$_2$O$_7$ in the misaligned sample ($\Delta\phi = 4.7$$^\circ$). Tetrahedra are numbered for an easy comparison with Fig.~\ref{f:CrystalStructure}. Such a spin-ice-like state is realized only for the misaligned fields of about 1.5\,T.}
\end{figure}

Figure~\ref{f:MagnAngle} shows the angles $\phi$ and $\theta$ of the Tb magnetic moments of Tb$_2$Ti$_2$O$_7$ and typical moment configurations for the $\bm{k} = \mathbf{0}$ magnetic structures are shown in Figs.~\ref{f:MagnStrA} 
%
and~\ref{f:MagnStrB}.
%
In all figures, quantities related to $\alpha$ and $\beta$ moments are shown by different colors, using the same conventions for clarity. The angles $\phi$ and $\theta$ refer to the [100] and [001] axes of the cubic unit cell, respectively, the field $\bm{H}$ is along the [110] axis.

Figure~\ref{f:MagnAngle} (left panel) shows the case of well-aligned fields ($\Delta\phi$ = 0.4$^\circ$ and 1.5$^\circ$). As seen from the $\theta(H)$ dependence, when the field increases, the Tb-$\alpha$ moments slowly rotate from their local [111] easy axis towards the field direction [110] lying in the (001) plane. Consequently, the antiferromagnetic component of the Tb-$\alpha$ moments along [001] becomes smaller. We note that at 12\,T the $\alpha$ moments still deviate, by about 12$^\circ$, from the field direction. As for the Tb-$\beta$ moments, we found that for a well-aligned field, they are collinear to the field. Surprisingly, they flip from a direction opposite to $\bm{H}$ ($A$ state, Fig.~\ref{f:MagnAngle}) to a direction parallel to $\bm{H}$ ($B$ state in Fig.~\ref{f:MagnAngle}), as shown by an abrupt change in the $\phi$ angle. This flip of the Tb-$\beta$ moments from an antiparallel to a parallel configuration occurs at a critical field $H_\mathrm{sf}$ of 1.5(5)\,T. Two typical spin structures below and above the critical field are drawn in Fig.~\ref{f:MagnStrA}.

Figure~\ref{f:MagnAngle} (right panel) shows the same quantities for fields misaligned (by $\Delta\phi$ = 4.0$^\circ$ and 4.7$^\circ$) toward the [100] direction. The comparison between the left and right panels clearly shows that the Tb-$\alpha$ moments are almost independent of the field misalignment. By contrast, a significant change occurs for the $\beta$ chains. In the samples misaligned by 4.0$^\circ$ and 4.7$^\circ$  the field variations in the $\phi$ angle are much smoother showing that the Tb-$\beta$ moments rotate from their local [111] easy axis toward the (001) plane when the field increases. In such cases, a spin-ice-like configuration of the magnetic moments is stabilized in fields of about 1.5\,T. This case corresponds to the spin-ice local order previously observed in Ref.~\onlinecite{prl.101.196402.2008} in the Tb$_2$Ti$_2$O$_7$ sample with a misalignment of 4$^\circ$ in the (001) plane, using unconstrained refinements. The spin-ice-like configuration is illustrated in Fig.~\ref{f:MagnStrB}. We notice that the field region where the spin-ice local order is stabilized in the misaligned samples roughly coincides with the critical field $H_\mathrm{sf}$ where the flip of the Tb-$\beta$ moments occurs in the well-aligned samples.

\begin{figure*}
\includegraphics[width=1.0\textwidth]{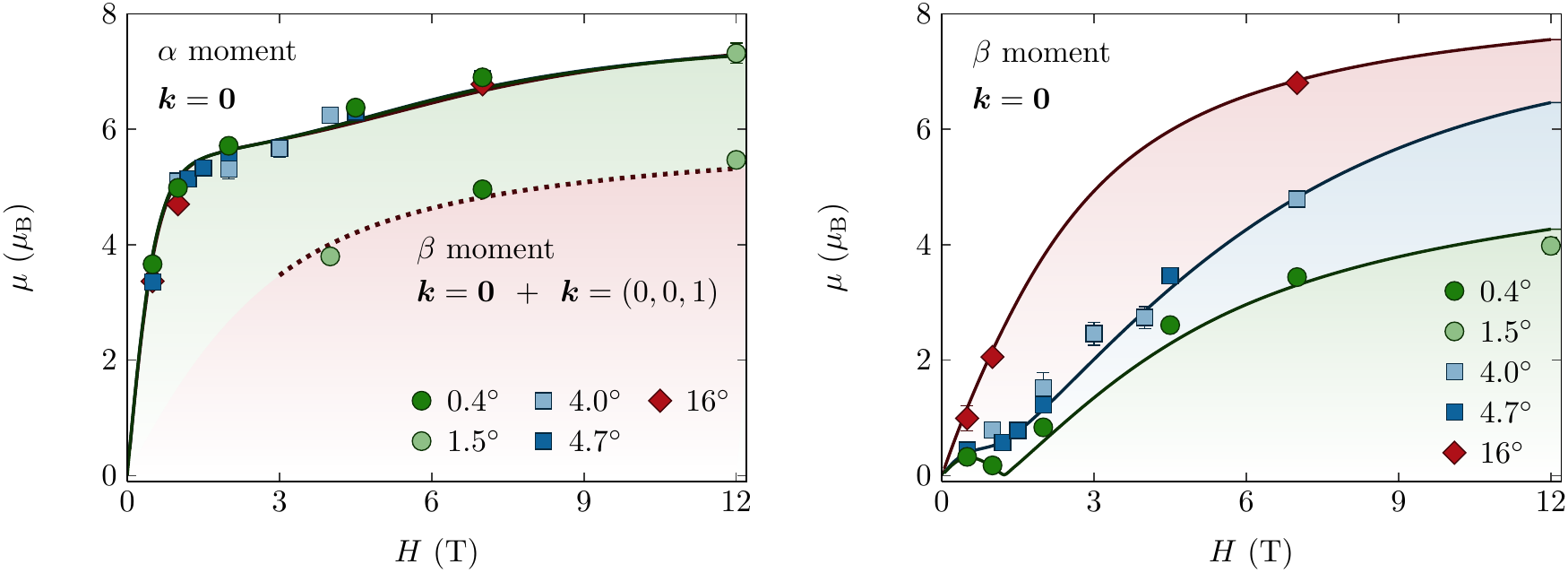}
\caption{\label{f:MagnMom}(Color online) Tb$_2$Ti$_2$O$_7$ $\bm{k} = \mathbf{0}$ structure at 1.6\,K. Field dependences of the Tb moment values for different field misalignments ($\Delta\phi$) with respect to the [110] direction. Error bars are smaller than the symbol size if not given. Left: Tb-$\alpha$ moments. Right: Tb-$\beta$ moments. The solid lines are calculations with the model described in Sec.~\ref{s:Model}. The following parameters were used for the calculations: $\lambda_\perp = -0.42$\,T/$\mu_\mathrm{B}$, $\lambda_\parallel  = -0.06$\,T/$\mu_\mathrm{B}$ and $\Delta\phi$ of 14$^\circ$, 3$^\circ$ and 0$^\circ$ from top to bottom (right). The total $\beta$ moment calculated as a vector sum of the $\bm{k} = \mathbf{0}$ and $\bm{k} = (0,0,1)$ moments is also shown at the left. The dotted line is a guide to the eyes.}
\end{figure*}

Another important information is given by the moment values (Fig.~\ref{f:MagnMom}).
%
In the $\bm{k} = \mathbf{0}$ structure, the $\alpha$ moments are practically independent of the field misalignment (Fig.~\ref{f:MagnMom}, left panel) but the $\beta$ moments strongly increase with the misalignment (Fig.~\ref{f:MagnMom}, right panel). Moreover, when the field is well aligned ($\Delta\phi = 0.4$$^\circ$), the $\beta$ moments show a minimum versus the field, decreasing to zero at $H_\mathrm{sf}$=1.5(5)\,T. This means that the spin flip occurs through a ``melting'' of the $\beta$ moments. Namely, the $\beta$ moments become paramagnetic at $H_\mathrm{sf}$ because the applied field exactly compensates the exchange field. Another example of spin ``melting'' is provided by the frustrated $XY$ model~\cite{prl.56.1074.1986}. The vanishing of the $\beta$ moments disappears together with the flip of the spins when the misalignment increases. One observes a small plateau for a misalignment of 4$^\circ$, then a strong and monotonous increase when the field is strongly misaligned by 16$^\circ$. In such a case, the magnitude and field dependence of the $\beta$ moments become comparable to those of the $\alpha$ ones, reflecting a gradual evolution from spin-ice-like to field-oriented structures with increasing field.

The influence of a field misalignement with respect to [110] can be qualitatively understood by its decomposition into a large [110] component along the $\alpha$ chains and a smaller perpendicular component, as suggested in Ref.~\onlinecite{prb.79.014408.2009} for Ho$_2$Ti$_2$O$_7$. In this case, the perpendicular component is along $[1\bar{1}0]$, namely, parallel to the $\beta$ chains, and tilts the $\beta$ moments. As to the $\alpha$ moments, a small misalignment should have little effect since the field component along [110] decreases very slightly. A quantitative explanation  of the field dependence of the $\bm{k} = \mathbf{0}$ structure is given by the model described below.

\subsection{\label{s:Model}\texorpdfstring{Model calculation for the $\mathbf{k=0}$ structure}{Model calculation for the k=0 structure}}

We implemented a model, based on the molecular field approximation, in order to reproduce the evolution with the field of the magnetic structure with $\bm{k} = \mathbf{0}$. This model performs mean-field self-consistent calculations and uses as ingredients the crystal field parameters of Tb$^{3+}$ in Tb$_2$Ti$_2$O$_7$ as in Refs.~\onlinecite{prb.76.184436.2007,prl.103.056402.2009}, and anisotropic two-ion exchange of the type:
%
\begin{equation}
\label{ech}
\mathcal{H}_\mathrm{ex} = -\mathcal{J}_\parallel\, S_{1Z}\, S_{2Z} - \mathcal{J}_\perp (S_{1X}\, S_{2X} + S_{1Y}\, S_{2Y}),
\end{equation}
where $\mathcal{J}_\parallel$ and $\mathcal{J}_\perp$ are the components of the exchange tensor in the local frame with trigonal symmetry.

We consider each ion in a tetrahedron to be exchange-coupled to its six nearest neighbors, resulting in an anisotropic molecular field tensor $\widetilde\lambda$ such that the molecular field acting on this ion writes:
%
\begin{equation}
\label{hmol}
\bm{H}_\mathrm{mol} = \frac{1}{6}\, \widetilde\lambda\, \sum_{k=1}^6 \bm{m}_k,
\end{equation}
where the sum runs over the six nearest neighbors.

The relationship between the components of the $\widetilde\lambda$ and $\mathcal{\widetilde J}$ tensors is:
%
\begin{equation}
\label{tens}
\lambda_i = 6 \mathcal{J}_i \left( \frac{g_\mathrm{J}-1}{g_\mathrm{J}} \right)^2 \frac{1}{\mu_\mathrm{B}^2},
\end{equation}
where $g_\mathrm{J}=3/2$ for Tb$^{3+}$.

A self-consistent treatment involving the four Tb moments, each with its three components, is performed, the only parameters being thus the two components of the $\widetilde\lambda$ tensor. This type of calculation is expected to describe only field induced magnetic structures having $\bm{k} = \mathbf{0}$ since only the moments on a single tetrahedron are considered. The dipole-dipole field is not included in the calculation.

For perfect alignment of the field with [110], the model is able to reproduce quite well all the experimental data (see solid lines in Figs.~\ref{f:MagnAngle} and~\ref{f:MagnMom}) with an anisotropic antiferromagnetic $\widetilde\lambda$ tensor. The spin-flip field $H_\mathrm{sf}$ of the $\beta$ moments, in particular, is a rapidly increasing function of the transverse component $\lambda_\perp$ and allows it to be determined with precision: $\lambda_\perp = -0.48(12)$\,T/$\mu_\mathrm{B}$, yielding $H_\mathrm{sf} = 1.5(5)$\,T. As to the longitudinal component $\lambda_\parallel$, we found that the isotropic exchange model with $\lambda_\perp$ = $\lambda_\parallel$ does not match to the experiment. An agreement can be obtained only above a certain limit: $\lambda_\parallel > -0.27$\,T/$\mu_\mathrm{B}$. In fact the calculated magnetic structures show very weak dependence on $\lambda_\parallel$, but only above this value ($-0.27$\,T/$\mu_\mathrm{B}$) the calculated low-field magnetic structure match the experiment.

The $\lambda_\perp$ value obtained here is about twice smaller than that ($-1$\,T/$\mu_\mathrm{B}$) derived in Ref.~\onlinecite{prl.103.056402.2009} from the thermal dependence of the transverse susceptibility $\chi_\perp(T)$. This can be due to the fact that the present model, involving four magnetic sub-lattices with six-neighbor exchange, is much more appropriate at low temperatures than the one-sublattice calculation performed in Ref.~\onlinecite{prl.103.056402.2009}. Let us mention that calculation of $\chi_\perp(T)$ with the present model (not shown) yields very good agreement with the data using $\lambda_\perp=-0.42(8)$\,T/$\mu_\mathrm{B}$, very close to the value derived above. This corresponds to a transverse exchange integral: $\mathcal{J}_\perp=-0.48$\,K.

For the case of a misalignment of the magnetic field with respect to [110], the model also reproduces quite well the data (see solid lines in Figs.~\ref{f:MagnAngle} and~\ref{f:MagnMom}) but the assessed misalignment angle in the (100) plane must be taken lower than the measured angle by about 0.5--1.5$^\circ$. We notice that the results depend slightly on the sign of the misalignment, either $\phi < 45$$^\circ$ or $\phi > 45$$^\circ$.

\subsection{\label{s:SymmMagn2}\texorpdfstring{Symmetry analysis and refinement of the $\mathbf{k=(0,0,1)}$ structures}{Symmetry analysis and refinement of the k=(0,0,1) structures}}

The symmetry analysis of the $\bm{k}=(0,0,1)$ structure was performed in the space group $I4_1/amd$ as in the case of the $\bm{k} = \mathbf{0}$ magnetic structure. In the same way, we find four two-dimensional irreducible representations $\Gamma'_i$ ($i=1$--4 according to Kovalev~\cite{book.kovalev.1965} notation) associated with $\bm{k} = (0,0,1)$, $\Gamma'_\mathrm{Tb} = 1\,\Gamma'^{(2)}_{1} + 2\,\Gamma'^{(2)}_{2} + 2\,\Gamma'^{(2)}_{3} + 1\,\Gamma'^{(2)}_{4}$. Hereafter, the symbol~$'$ denotes the parameters which correspond to the $\bm{k}=(0,0,1)$ structure. Among the representations, only $\Gamma'_3$ (see Table~\ref{t:IrrRepTb2}) is in agreement with our experimental data. As seen from Table~\ref{t:IrrRepTb2} four fitting parameters are needed to describe the $\Gamma'_3$ representation: two of them ($U'_y$ and $U'_z$) describe the $\alpha$ chains and the remaining two ($V'_x$ and $V'_z$) describe the $\beta$ chains.

\begin{table}
\caption{\label{t:IrrRepTb2}Irreducible representation $\Gamma'_{3}$ of Tb$_2$Ti$_2$O$_7$ (space group $I4_1/amd$) associated with $\bm{k}=(0,0,1)$. Basis vectors projected from a general vector $\bm{M}$ with the components $M_x$, $M_y$, and $M_z$ at the Tb $8d$ sites. The Tb positions are defined in Table~\ref{t:IrrRepTb1}.}
\begin{ruledtabular}
\begin{tabular}{llrrr}
Irr. Rep.     &Atom          &\multicolumn{1}{c}{$M_x$} &\multicolumn{1}{c}{$M_y$} &\multicolumn{1}{c}{$M_z$} \\
\colrule
$\Gamma'_{3}$ &Tb-$\alpha_1$ &\multicolumn{1}{c}{0}     &$U'_y$                    & $U'_z$                   \\
              &Tb-$\alpha_2$ &\multicolumn{1}{c}{0}     &$U'_y$                    &$-U'_z$                   \\
              &Tb-$\beta_1$  &$V'_x$                    &\multicolumn{1}{c}{0}     &$-V'_z$                   \\
              &Tb-$\beta_2$  &$V'_x$                    &\multicolumn{1}{c}{0}     & $V'_z$                   \\
\end{tabular}
\end{ruledtabular}
\end{table}

\begin{figure}
\includegraphics[width=1.0\columnwidth]{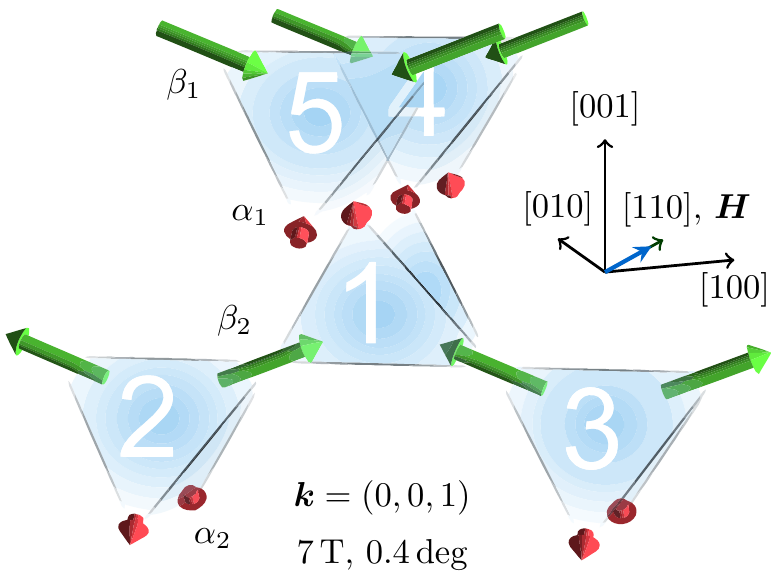}\vspace{2.5ex}
\caption{\label{f:MagnStrC}(Color online) Field-induced magnetic structure with $\bm{k} = (0,0,1)$ of Tb$_2$Ti$_2$O$_7$ in the well-aligned sample ($\Delta\phi = 0.4$$^\circ$). Tetrahedra are numbered for an easy comparison with Fig.~\ref{f:CrystalStructure}.}
\end{figure}

The $\bm{k}=(0,0,1)$ magnetic structure was investigated in the well-aligned samples under an applied field up to 12\,T. In the case of $\Delta\phi = 0.4$$^\circ$, a total of 196 pure magnetic reflections with $\ensuremath{\sin\theta/\lambda} < 0.55$\,\AA$^{-1}$ were measured at 1.6\,K and 7\,T. They were used to refine four Tb magnetic moment components ($R_\mathrm{F} = 0.046$) taking into account the symmetry constraints of $\Gamma'_3$. The magnetic structure resulting from the refinement is illustrated in Fig.~\ref{f:MagnStrC}. We find that the Tb-$\alpha$ moments are very small, typically 0.2--0.6\,$\mu_\mathrm{B}$, as already quoted in Ref.~\onlinecite{prl.101.196402.2008}, so that the magnetic order mostly concerns the Tb-$\beta$ moments. The $\bm{k}=(0,0,1)$ propagation vector yields an AFM coupling of the $\beta$ chains. Within one chain, the $\beta$ moments order antiferromagnetically along the chains. We notice that such AFM order along the $\beta$ chains contrasts with the $\bm{k}=X$ magnetic structure observed in Ho$_2$Ti$_2$O$_7$ spin ice \cite{prl.79.2554.1997}. In Ho$_2$Ti$_2$O$_7$, the $\beta$ chains also couple antiferromagnetically but the moments within one chain are ferromagnetically coupled along the chain.

\subsection{\texorpdfstring{Resulting magnetic structure of Tb$_2$Ti$_2$O$_7$}{Resulting magnetic structure of Tb2Ti2O7}}

\begin{figure}
\includegraphics[width=1.0\columnwidth]{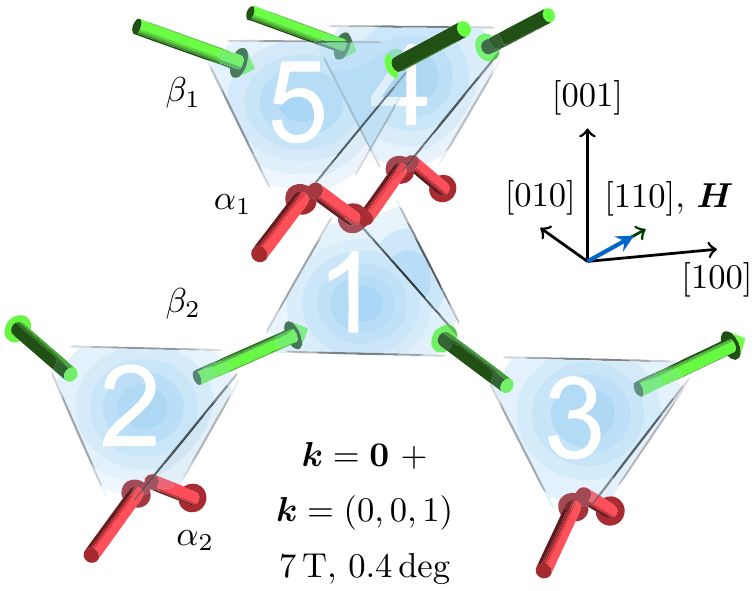}
\caption{\label{f:MagnStrD}(Color online) Field-induced magnetic structure of Tb$_2$Ti$_2$O$_7$ in the well-aligned sample ($\Delta\phi = 0.4$$^\circ$). A superposition of $\bm{k} = \mathbf{0}$ and $\bm{k} = (0,0,1)$ is shown. Tetrahedra are numbered for an easier comparison with Fig.~\ref{f:CrystalStructure}.}
\end{figure}

First we recall that the $\bm{k} = (0,0,1)$ structure is stabilized only at low temperatures, in high and well-aligned fields (typically above 2\,T and for a misalignment below 4.5$^\circ$). In all other cases, only the $\bm{k} = \mathbf{0}$ structure is observed. In the ($H$, $T$, $\Delta\phi$) range where the two structures coexist, one can easily reconstruct the full magnetic structure of Tb$_2$Ti$_2$O$_7$ from the refined $\bm{k} = \mathbf{0}$ and $\bm{k} = (0,0,1)$ structures, as a coherent superposition of the magnetic modes involved in $\Gamma_{9}$ and $\Gamma'_3$ representations. This resulting magnetic structure of Tb$_2$Ti$_2$O$_7$ at 7\,T is illustrated in Fig.~\ref{f:MagnStrD}.

The field dependences of the Tb magnetic moments resulting from this superposition are shown in Fig.~\ref{f:MagnMom}. The $\alpha$ moments which are involved in the $\bm{k} = \mathbf{0}$ structure only increase with the field and reach 7\,$\mu_\mathrm{B}$ at 12\,T, whatever the field alignment. By contrast, the $\beta$ moments, which share between the two structures, strongly depend on it. In well-orientated samples at high fields, a $\beta$-moment component of about 4\,$\mu_\mathrm{B}$ belongs to the $\bm{k} = \mathbf{0}$ structure and another component of about 4\,$\mu_\mathrm{B}$ goes to the $\bm{k} = (0,0,1)$ structure. However, the total $\beta$ moment calculated as the vector sum of the $\bm{k} = \mathbf{0}$ and $\bm{k} = (0,0,1)$ moments (Fig.~\ref{f:MagnMom}, left panel) reaches only about 5.5\,$\mu_\mathrm{B}$, thus remaining smaller than the $\alpha$ moment. When the misalignment is above 4.5$^\circ$ and the $\bm{k} = (0,0,1)$ structure is suppressed, the $\beta$ moments in the $\bm{k} = \mathbf{0}$ structure reach the value of the $\alpha$ moments of about 7\,$\mu_\mathrm{B}$ in high fields. We notice that this 7\,$\mu_\mathrm{B}$ value is above the value of 5.1\,$\mu_\mathrm{B}$ found at $H = 0$\,T for the fluctuating Tb moments~\cite{prb.76.184436.2007}. This is explained by the mixing of the low-energy crystal field levels by the applied field.

\begin{figure}
\includegraphics[width=1.0\columnwidth]{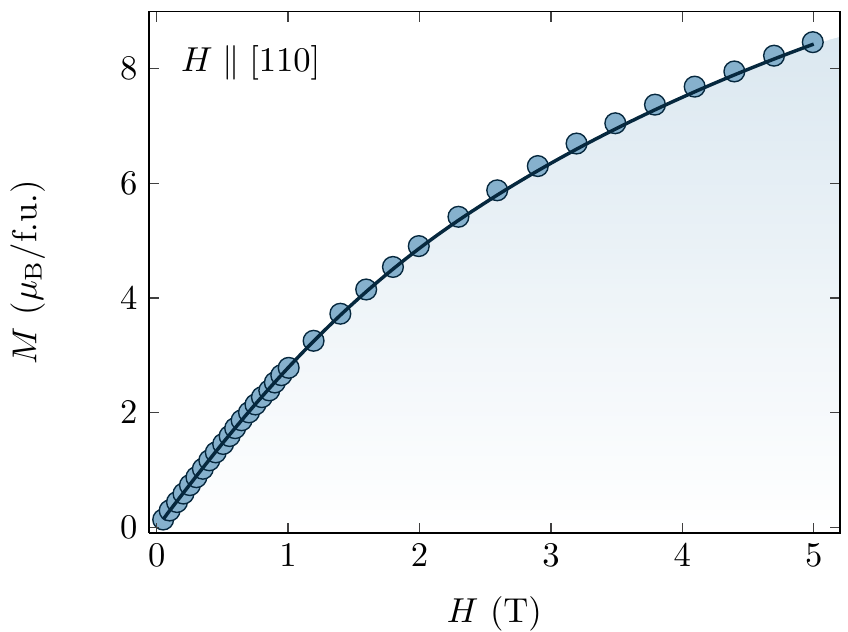}
\caption{\label{f:MvsH}(Color online) Magnetization $M$ of Tb$_2$Ti$_2$O$_7$ measured at 5\,K versus the magnetic field $H$ applied along the [110] axis according to Ref.~\onlinecite{jpsj.71.599.2002} (circles). The solid line is a calculation with the model described in Sec.~\ref{s:Model} using the molecular field tensor deduced from the present measurements.}
\end{figure}

From our measurements and the mean field calculation described above, we can also calculate the magnetization and compare it to its determination by bulk magnetic measurements~\cite{jpsj.71.599.2002}. As shown in Fig.~\ref{f:MvsH}, both quantities are in  perfect agreement, which is an important check of the robustness and consistency of our analysis and which ensures that all the experimental corrections have been done properly. One also notices that misorientation effects become negligible here since the bulk magnetization is almost independent of the crystal orientation~\cite{jpsj.71.599.2002}.

\section{Conclusion}

To summarize, we studied the field-induced magnetic structures in Tb$_2$Ti$_2$O$_7$ under  magnetic field close to the [110] axis, by varying the field alignment in a systematic way. We observed different types of non-collinear structures which are extremely sensitive to the field alignment. This sensitivity mostly concerns the moments on the $\beta$ chains whose local anisotropy axis is perpendicular to the field. In low fields ($H < 2$\,T), when only the $\bm{k} = \mathbf{0}$ structure is stabilized, spin-ice local orders are favored for a misalignment of a few degrees. Local orders where the Tb-$\beta$ moments flip along the field direction and melt at a compensation field, occur when the sample is very well aligned (within 0.4$^\circ$).

We note that the mechanism in the flip of the Tb-$\beta$ moments in Tb$_2$Ti$_2$O$_7$ is considerably different from the spin-flop and spin-flip transitions in conventional antiferromagnets such as MnF$_2$ and FeCl$_2$, respectively~\cite{jap.32.S61.1961,pr.164.866.1967}. In these materials, the moment reorientation occurs as a result of the competition between anisotropy, exchange and Zeeman energy, it is accompanied by a symmetry breaking and it is characterized by a moment conservation. By contrast, no symmetry breaking occurs in Tb$_2$Ti$_2$O$_7$ at the critical field $H_\mathrm{sf}$ since both magnetic structures above and below $H_\mathrm{sf}$ belong to the same irreducible representation. Moreover, the magnetic moment of the Tb-$\beta$ chains gradually decreases until the applied field exactly compensates the exchange field for $H = H_\mathrm{sf}$. At this point, the $\beta$ moments become paramagnetic, which opens the possibility to observe new kinds of field induced excitations in Tb$_2$Ti$_2$O$_7$. The parent compound Er$_2$Ti$_2$O$_7$ recently studied offers an intermediate case, with a decrease of the Er moment values at the spin-flip transition~\cite{prb.82.104431.2010}.

The field behavior of the $\bm{k} = \mathbf{0}$ structure is well explained by a molecular field model with anisotropic exchange. In high fields ($H > 2$\,T), the antiferromagnetic structure with $\bm{k} = (0,0,1)$ propagation vector, which involves mostly the $\beta$ moments, is favored in well-aligned samples and rapidly disappears in the misaligned samples. In this structure, the $\beta$ moments are antiferromagnetically coupled along the chains, in contrast with Ho$_2$Ti$_2$O$_7$ spin ice. The high sensitivity of the magnetic orders to the field alignment results from the combination of the crystal-field anisotropy and pyrochlore geometry, which tunes the coupling of the $\beta$ chains to the applied field.

We performed a full symmetry analysis of the field induced magnetic structures in the pyrochlore lattice. The field evolution of the magnetic structure in Tb$_2$Ti$_2$O$_7$ is successfully described using single irreducible representations of the $I4_1/amd$ space group, both for the $\bm{k} = \mathbf{0}$ and $\bm{k} = (0,0,1)$ structures. The main advantage of the symmetry analysis is a reduction of the fitted parameters needed in the refinement, six parameters for the $\bm{k} = \mathbf{0}$ structure and only four for the $\bm{k} = (0,0,1)$ one, instead of twelve in the unconstrained refinement. Similar approaches may be used to analyze the field-induced magnetic structures of other pyrochlores.

\emph{Note added}: Recently, we were made aware of the study from another group~\cite{prb.82.100401.2010} on the same subject.

\begin{acknowledgments}
We thank E.~Ressouche  for his help during the experiment at the Institut Laue Langevin. H.~C. acknowledges the support of Le Triangle de la Physique.
\end{acknowledgments}

\bibliography{biblio}

\end{document}